\documentclass[namedreferences]{SolarPhysics}
\usepackage[optionalrh]{spr-sola-addons} 
\usepackage{graphicx}                    
\usepackage{color}                       
\usepackage{url}                         
\usepackage{lscape}
\usepackage{rotating}

\usepackage{solaheader} 
\newcommand{\solareceiveddate}{13 May 2013} 
\newcommand{\solaaccepteddate}{11 July 2013}

\begin{document}

\begin{article}

\begin{opening}

\title{Properties of Magnetic Neutral Line Gradients and Formation of Filaments}
\author{Nina V.~\surname{Karachik}\sep 
Alexei A. \surname{Pevtsov}}

%

\runningauthor{Karachik and Pevtsov} 
\runningtitle{Properties of Magnetic Neutral Line Gradients and Formation of Filaments}


\institute{National Solar Observatory, Sunspot, NM 88349, U.S.A. \\
email: \url{nkarachi@nso.edu;apevtsov@nso.edu}}

\begin{abstract}
                                          
We investigate gradient of magnetic field across neutral lines (NLs),
and compare their properties for NLs with and without the chromospheric
filaments.  Our results show that there is a range of preferred magnetic field
gradients where the filament formation is enhanced. On the other hand, 
a horizontal gradient of magnetic field across a NL alone does not 
appear to be a single factor that determines if a filament will form 
(or not) in a given location.
\end{abstract}


\end{opening}

\section{Introduction}
                                                 
Chromospheric filaments (prominences at the limb) are one of the major
features characterizing the solar activity in low solar atmosphere
(chromosphere and low corona). Filament eruptions are at the core of
many solar-terrestrial effects or Space Weather. The filaments are
formed along magnetic polarity inversion lines, but only a minority of
these neutral lines (NLs) has filaments above them. Several studies
were made trying to find the necessary conditions for filament
formation, but still there is no clear understanding of this
\cite{Mackay2010}. It was found by previous studies that filaments form
only in filament channels (for review, see \opencite{Gaizauskas1998}),
which overlay NLs.  Some studies have indicated that the convergence of
magnetic flux leading to either flux cancellation or reconnection is
required for filament formation (see \opencite{Mackay2008};
\opencite{Martin1998}; \opencite{Martin1990}, and references therein).
Recently, \inlinecite{Martin2012} suggested that formation of filament
and filament channels is part of development of large-scale chiral
systems. In our present study, we concentrate only on properties of
neutral lines, and do not consider other aspects of filament formation
(such as, for example, role of helicity).

\inlinecite{Gaizauskas2001} have shown that one important condition for
the formation of a filament channel is a strong horizontal component of
magnetic field aligned with the magnetic neutral line.  Thus, one can
ask if the properties (and evolution) of magnetic NLs could be
sufficient to identify potential location of filament formation?
Indeed, several studies  (\textit{e.g.}, \opencite{Maksimov1995};
\opencite{Nagabhushana1990}) have suggested that the gradient of
magnetic field transverse to the neutral line may play a significant
role in filament formation. According to \inlinecite{Demoulin1998} one
of the necessary conditions for the filament formation is a low
gradient of magnetic field transverse to NL.

Here we conduct a statistical study of horizontal gradients of magnetic
field across neutral lines, and investigate how these properties differ
for NLs with filaments and without them.  We also investigate whether the
properties of neutral lines change with solar cycle, and whatever this change
can be the cause of variation in filaments length and total number with 
the solar cycle.

\section{Data and Data Analysis\label{sec:data}}

We employ daily H$\alpha$ full-disk images from  Mauna Loa Solar
Observatory (MLSO) and SOHO/MDI (\textit{Michelson Doppler Imager}) synoptic
maps for 1997-2010 years. For our main H$\alpha$ data set, we included
daily full-disk images corresponding to two complete  Carrington
rotations (CRs) per year with a time separation between selected CRs
about five months.  However, due to lack of data, only one rotation
was taken for the years 1997, 2003, and 2010; and for year 2000, we
selected three (non-consecutive) solar rotations.  In addition to MDI
synoptic maps (for same Carrington rotations as our H$\alpha$ data), we
also use the \textit{Synoptic Optical Long-term Investigations of the Sun}
(SOLIS, \opencite{Bala2011}) synoptic maps and the full-disk
magnetograms taken at Ca II 854.2 nm chromospheric line to study
properties of the chromospheric neutral lines.

Because of the statistical nature of our study, we automated the
identification of filaments (using H$\alpha$ full-disk images) and
neutral lines (using MDI synoptic maps). Several studies were made in
developing filament automatic detection (\textit{e.g.}, \opencite{Shih2003};
\opencite{Qu2005}; \opencite{Scholl2008}; \opencite{Joshi2010}).  Most
of them have an extensive algorithm of filament detection from an image
taken in a single wavelength or use multiple images taken in different
wavelengths. In our algorithm, we used both H$\alpha$ daily full-disk
images and magnetograms. This allowed us to significantly simplify the
process of filament detection by comparing the location of potential
filaments (dark features initially found on H$\alpha$ images) with
underlying magnetic field. This approach also helped to reduce
(computer) processing time.

The criterion for filament identification is an intensity threshold
which is defined as an average intensity over the full-disk image
minus  two standard deviations of the average. All pixels satisfying
the above criterion (but excluding pixels near the limb) are considered
as potential filaments. The final selection takes into account
existence of NL inside the filament contour. Magnetic neutral lines
were identified on synoptic maps after smoothing them by 50 x 50 pixels
running window.

\begin{figure}
\includegraphics[width=1.0\textwidth,clip=]{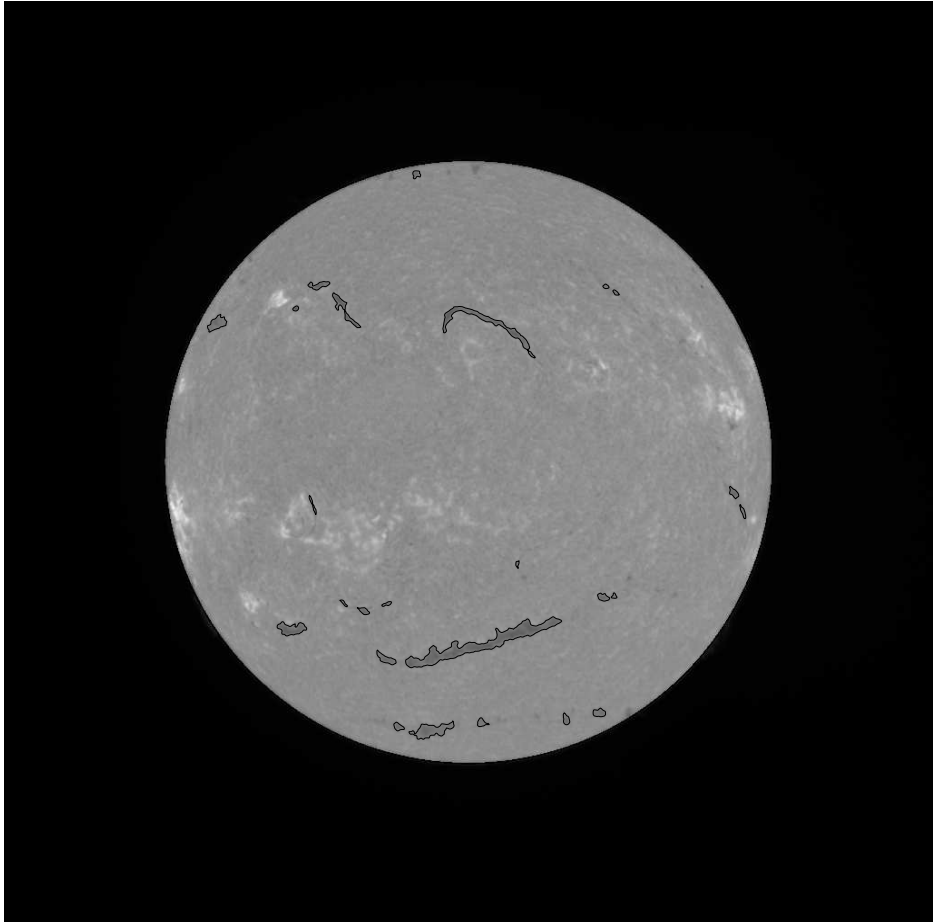}
\caption{H${\alpha}$  full-disk image with filament contours as identified 
by our automated procedure.
\label{fig1}}
\end{figure}
                                                 
Figure 1 shows an example of a H$\alpha$ full-disk image with filament
contours found by our automated procedure. It is clear that our
procedure  successfully identifies all major filaments. Small clumps
of filament material  could be missed especially in locations near the
solar limb. For example, while the procedure identifies one clump of
filament material near the North polar region  (near the top of solar
disk in Figure 1), it misses several other clumps in the same
location. As an additional test, for a few selected solar rotations,
we compared our filament identification with the data from the
BASS2000 Solar Survey Database (bass2000.obspm.fr).  Figure 2
demonstrates a very good agreement between filaments from BASS2000
data base (green), and those identified by our method (red).

\begin{figure}
\includegraphics[width=1.0\textwidth,clip=]{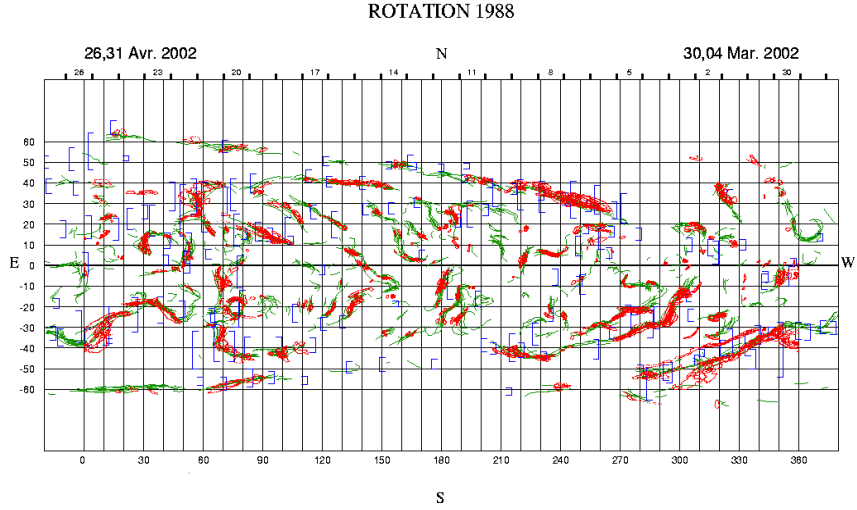}
\caption{Filaments from BASS2000 Solar Survey Database (green) and our 
identification method (red).
\label{fig2}}
\end{figure}

\begin{figure}
\includegraphics[width=1.0\textwidth,clip=]{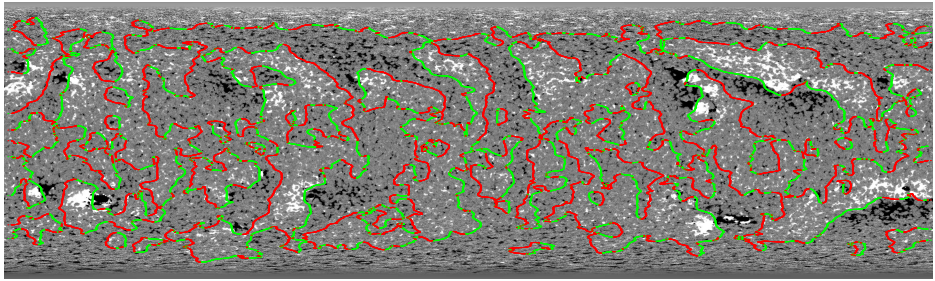}
\caption{Example of MDI synoptic map for the Carrington Rotation 1932 with 
neutral lines. Green color is used for type A lines (inside one bipolar 
region) and red is for type B lines (between two bipolar regions). 
Black/white corresponds to negative/positive polarities.
\label{fig3}}
\end{figure}
                                                 
Figure 3 is an MDI synoptic map with neutral lines drawn in different
colors. The colors are used to indicate the type of NL: green is for
the lines between magnetic polarities inside one active region, and
red is for the lines separating polarities of two different active
regions.  For the reasons of automation, the determination of a type
of NL (inter/intra- active region) is based on the polarity of fields
to the West and East of the neutral line. If the orientation of these
two polarities followed the Hale polarity rule, we called the neutral
line as intra-active region.  Otherwise, it was classified as
inter-region NL. Inevitably, this approach mis-identifies non-Hale
polarity active regions. But the fraction of such regions is very
minor, and thus, the effect on overall statistics is insignificant.

\begin{figure}
\includegraphics[width=1.0\textwidth,clip=]{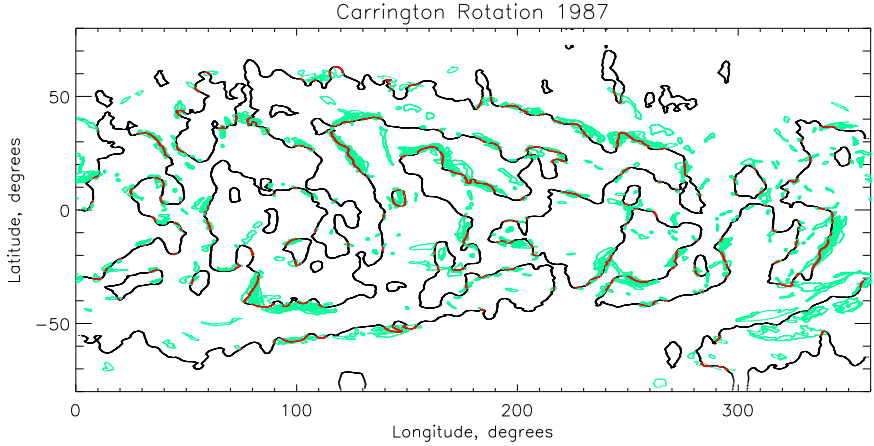}
\caption{Example of neutral lines (black) with filament contours (green). 
Parts of neutral lines that have filaments are marked by red color.
\label{fig4}}
\end{figure}
                                                           
Figure 4 provides example of neutral lines associated with the
chromospheric filaments as identified by our procedure. The portions
of neutral lines with filaments are marked by red color.

Although we analyze only two solar rotations per year, our data set of
filaments is representative of a complete data. For example, a
latitudinal distribution of filaments in our data set exhibits
well-known latitudinal drift (\textit{e.g.}, \opencite{Li2010}). 
During the rising phase of solar activity, the filaments distribution
(in our set) shows  a clear preference for high latitudes, and as
the cycle progresses, more filaments are found at lower latitudes
following the equatorward drift similar  to sunspot's  butterfly
diagram.

\section{Displacement of filaments with respect to neutral lines}
                                                 
While the chromospheric filaments are located above the magnetic
neutral lines, it is not uncommon to see a displacement between the
location of a NL (as measured in the photosphere) and the filament
body (as observed in H$\alpha$) in full-disk images. One can
contribute such a displacement to the fact that filaments are the
features located at certain heights above the photosphere and to their
projection to the image plane.

However, our analysis suggests that this displacement could not be
explained by a simple projection effect alone. To further explore this
displacement, we analyzed a number of filaments observed near the
central meridian. High-latitude filaments ($\vert$lat$\vert {\geq}$ 60
degrees) were excluded from this analysis.  We also compared the
change in filament-NL displacements for selected filaments near the
East and West limbs.  The displacements between the filament body and
the neutral line were present in different locations independent of
closeness to the limb or the central meridian. In some cases, only a
part of a filament was displaced relative to closest neutral line,
while the rest of the filament was perfectly aligned with NL.  Figure
5 (left panel) shows one such an example.

Since the filaments could be located at a certain height above the
photosphere, one could argue that  their location should correlate
better with the neutral lines of the  chromospheric fields (which are
formed higher in the atmosphere).  Contrary to this expectation, some
filaments show a larger displacement relative to closest chromospheric
neutral line as compared  with the photospheric NL (compare left and
right panels in Figure 5). Thus, the displacement of filament location
(as seeing in image plane of the photospheric and chromospheric
magnetograms) is clearly not due to a projection effect alone.

\begin{figure}
\hbox{
\includegraphics[width=0.45\textwidth,clip=]{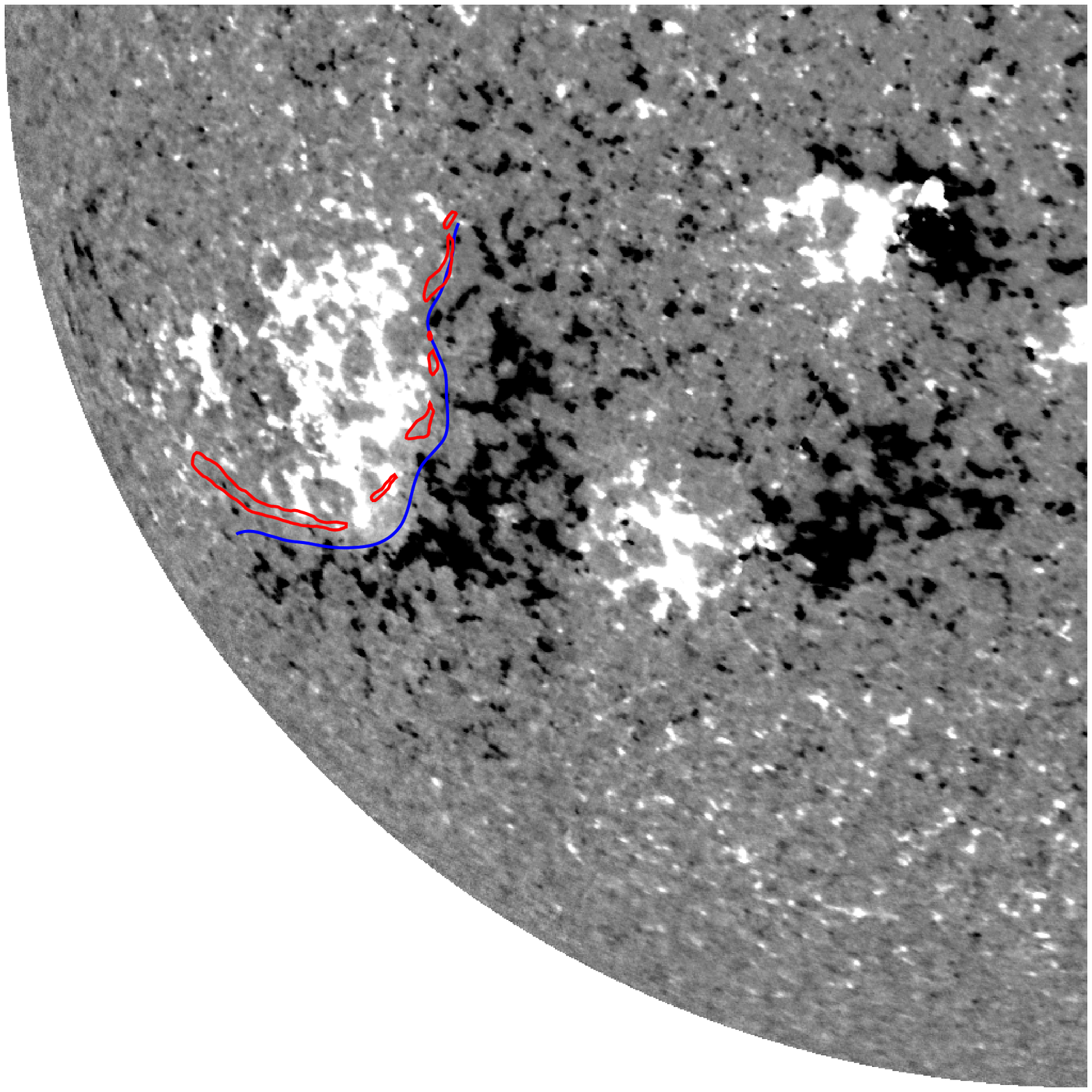}
\includegraphics[width=0.45\textwidth,clip=]{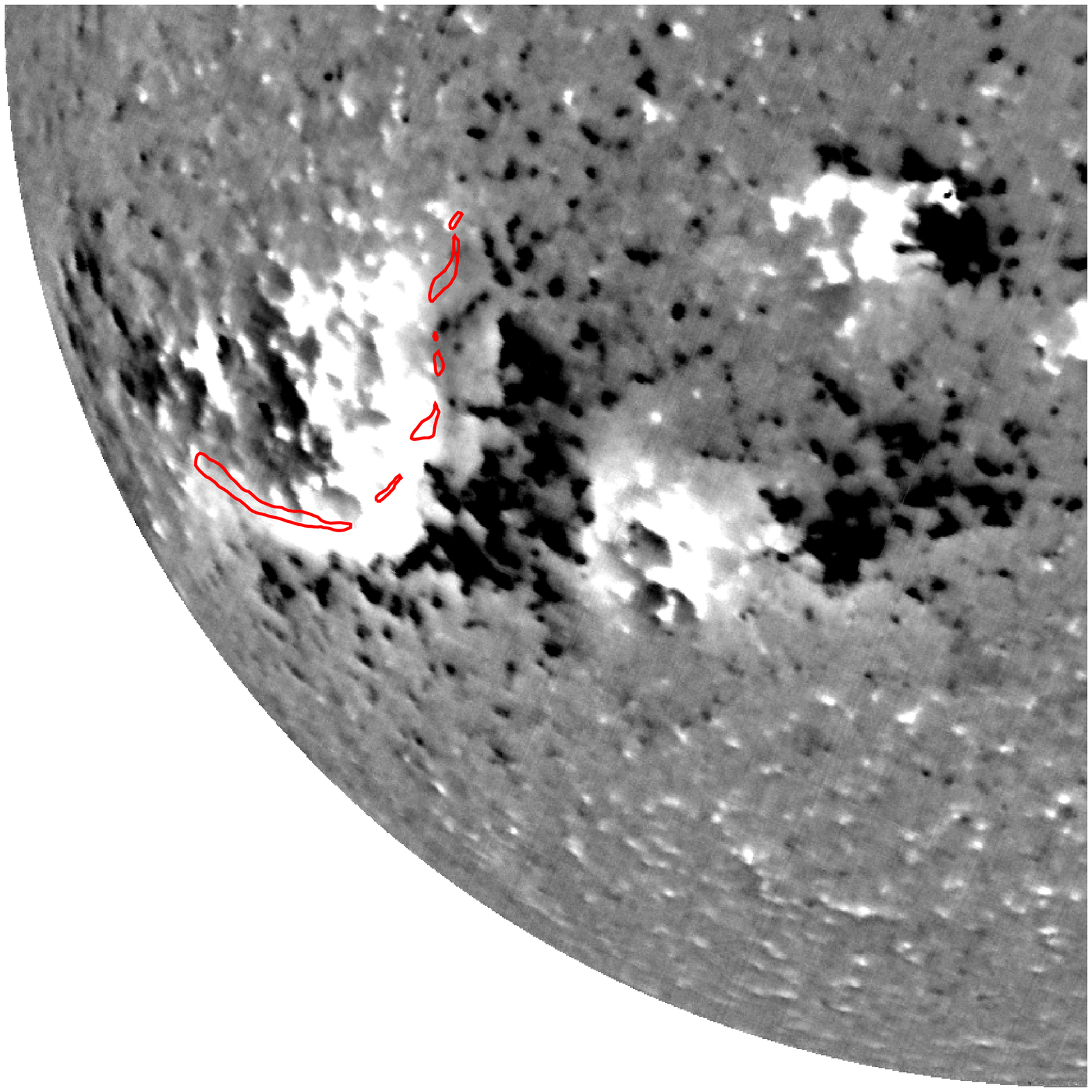}
}
\caption{Example of filament displacement with respect to neutral line on 
photospheric (left) and chromospheric (right) magnetogram from VSM on SOLIS. 
Blue line 
represents neutral line, red contour - a filament. Black/white corresponds 
to negative/positive polarities.
\label{fig5}}
\end{figure}
                            
One possible explanation for the displacements described above is that
the "neutral lines" are not lines but planes separating opposite
polarities. These "polarity inversion planes" are not always vertical,
but they could be inclined depending on flux systems they
separate. One can visualize this as a ribbon (perhaps, a current
sheet) ``waving'' around the isolated coronal flux systems. The
intersection of this polarity inversion plane with the photosphere
corresponds to the photospheric neutral line, and the filament can
form within this plane higher up.  If this interpretation is correct,
the displacement between the photospheric  NL and filament body would
be an interplay between the location (height) of  filament within the
polarity inversion plane and the tilt of the plane  relative to solar
surface.  The idea that (some) chromospheric filaments are sheet-like
structures is not new. We simply use it to demonstrate that a broadly
held belief that the filaments are located above magnetic neutral
lines should be taken with a clear understanding that in image plane
that location of filament bodies and the neutral lines could be
displaced from each other, and not only due to elevation of filament
above the photosphere, but also because of its height geometry.

In some cases of relatively weak (and fragmented) magnetic fields, the
exact location of the photospheric neutral line is very uncertain.  In
some of these cases, a continuous photospheric neutral line may not
even exist underneath of a filament. For example, filament segment in
the low-right corner of Figure 6 overlies patch of negative polarity;
there appears no clear  neutral line (of photospheric field) that
can be associated with this  filament segment.

\begin{figure}
\hbox{
\includegraphics[width=0.45\textwidth,clip=]{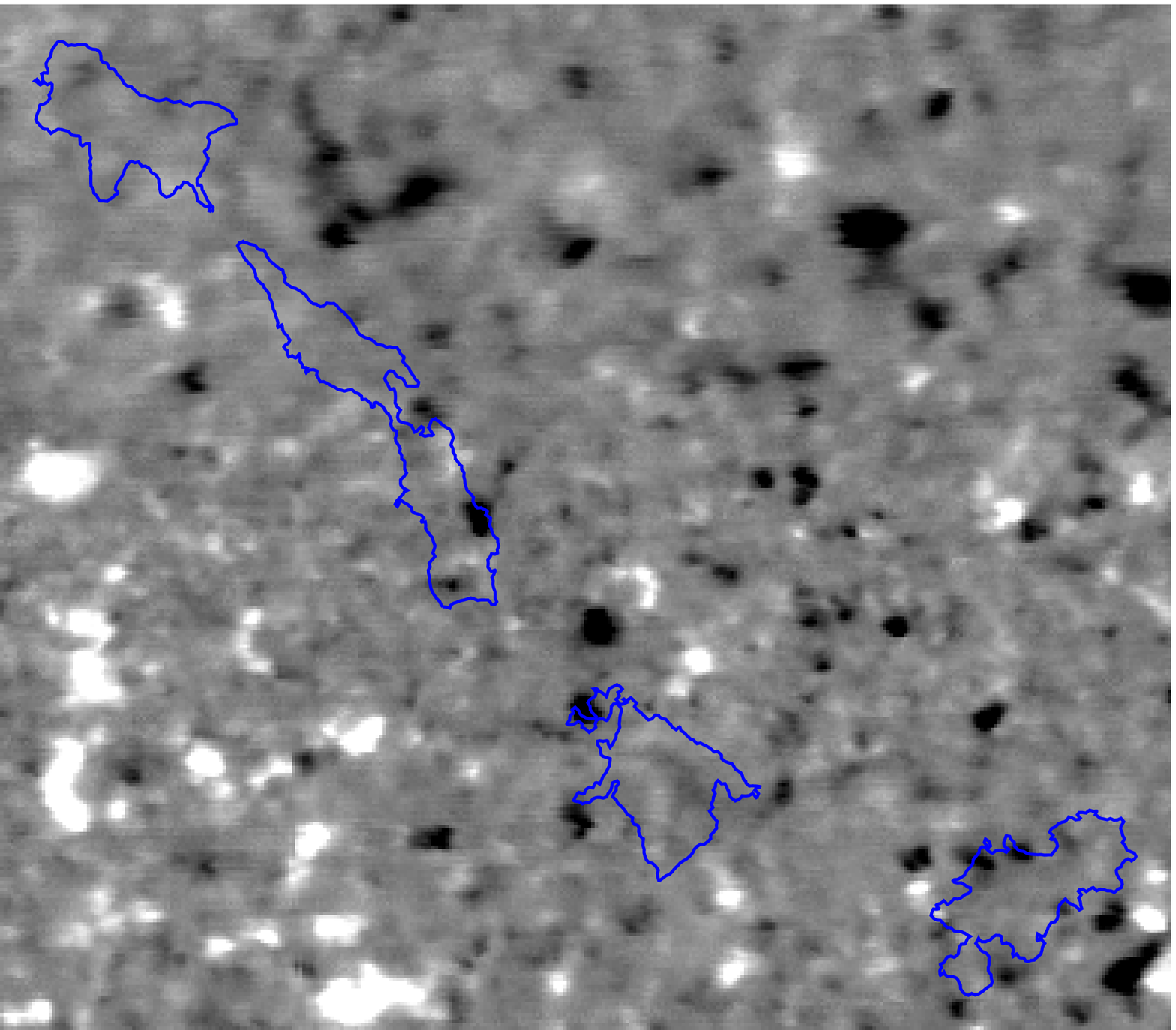}
\includegraphics[width=0.45\textwidth,clip=]{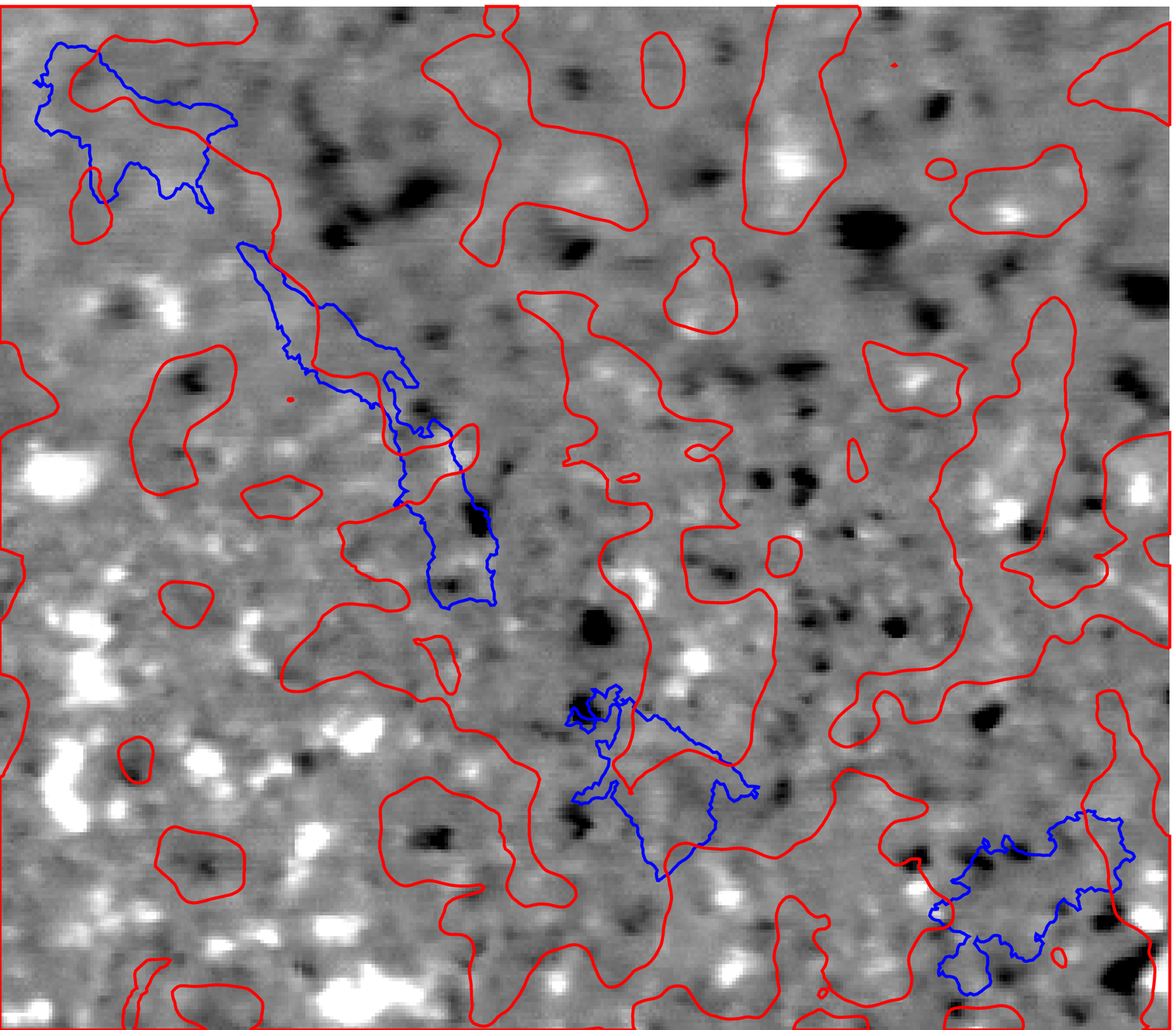}
}
\caption{Example of a filament contour (blue) displayed on magnetogram. Red 
color is used for neutral lines. Black/white corresponds to negative/positive 
polarities.
\label{fig6}}
\end{figure}

The inconsistencies between the location of filaments and the
photospheric/ chromospheric neutral lines described in this section
may  indicate that the filament location is governed by the coronal
magnetic  fields rather than the photospheric or even the
chromospheric fields.
                                                 
As a rule, we find a better agreement between the neutral lines and
filament positions when we use synoptic maps. We speculate that since
these maps are  based on an averaged magnetic field, they may better
represent the large-scale magnetic domains. Even though for filament
identification we employ daily full-disk images, we do average
filament positions over several days, and thus, any  projection
effects are minimized.

Based on the results of the analysis presented in this Section, for
our  study of magnetic properties of NLs we chose to use the  synoptic
magnetograms.
                                               
\begin{figure}
\includegraphics[width=1.0\textwidth,clip=]{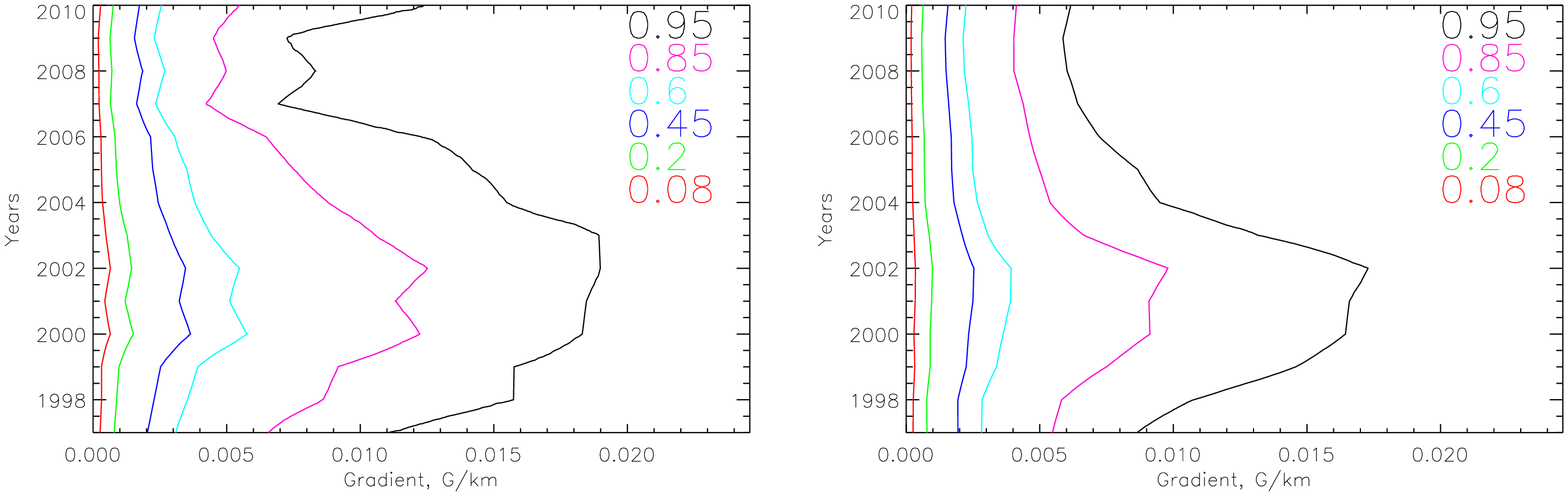}
\caption{Temporal variation of gradient distribution. Left - for NLs with 
filaments, right - for lines without filaments. Colored contours correspond 
to fraction of gradients below a certain value. For  instance, for year 1997 
95\%
of all NL without filaments have gradients below  0.0088 G km$^{-1}$ (right 
panel). 
\label{fig7}}
\end{figure}
                                                 
\section{Role of gradient of magnetic field\label{sec:results}}
                     
First, we investigate gradients of magnetic field across the neutral
line and their variation with level of solar activity. Figure 7 shows
temporal variation of gradient distribution for NLs with filaments
(left panel) and without them (right panel). The gradients were
computed at uniformly spaced intervals along magnetic neutral lines.
Then, each segment was identified either as having a filament or not
based on proximity of filament material to this part of neutral
line. We did not attempt any averaging of field gradients for
individual filaments.  Different colors are used to indicate the
fraction of NLs with a specific property. For example, contour labeled
0.95 means that 95\% of the NLs have gradients less or equal of
gradient shown as abscissa on Figure 7. For instance, for the year 1997,
95\%  of NLs with filaments (Figure 7, left) have gradients ${\le}$
0.011 G km$^{-1}$, while 95 \%  of NLs without filaments (Figure 7,
right) have gradients ${\le}$ 0.0088 G km$^{-1}$.

At the first glance, the distributions of gradients across NLs with
and without filaments are very similar. However, neutral lines with
filaments exhibit systematically stronger gradients as compared with
NLs without  filaments (compare Figure 7 left and right panels).
Filaments are observed above the NLs with gradients ranging from 6.6
$\times$ 10$^{-5}$ G km$^{-1}$ to 0.11 G km$^{-1}$. Only a small
fraction of filaments are formed above the NLs with gradients outside
that range. A fraction of the neutral lines with gradients smaller
than about 0.001 G km$^{-1}$ is about 20\% for NLs with or without
filaments, and it does not seem to vary with solar activity cycle. A
fraction of neutral lines with gradients higher than 0.005 G km$^{-1}$
shows a strong variation with solar activity cycle. By its amplitude,
NLs with and without filaments show comparable change between solar
minimum and maximum: for NLs with filaments, 95\% level decreases from
about 0.019 G km$^{-1}$ (year 2002) to about 0.007 G km$^{-1}$  (year
2009), while for NLs without filaments it changes from 0.017 G
km$^{-1}$ (year 2002) to 0.006 G km$^{-1}$ (year 2009). On the other
hand, the changes for NLs with filaments occur over a longer period of
time (\textit{i.e.}, peak corresponding to 95\% level on Figure 7 left
panel is much "broader" as compared with Figure 7, right panel). In
2002, about 50\% (between 0.95 and 0.45 contours in Figure 7) NLs with
filaments fall within 0.019-0.0035 G km$^{-1}$. In the same year, 35\%
of all filaments (between 0.95 and 0.45 contours in Figure 7) formed
above NLs within 0.019-0.005 G km$^{-1}$ range of gradients. In
comparison,  50\% of NLs without filaments fell within 0.017-0.0025 G
km$^{-1}$, and 35\%  of NLs (without filaments) were in 0.017-0.004 G
km$^{-1}$ range.  In 2004, 50\% (0.95-0.45 levels in Figure 7) of
filaments developed above  NLs with 0.0155-0.0022 G km$^{-1}$ magnetic
field gradients, while 50\% NLs without filaments fall within
0.0095-0.0018 G km$^{-1}$ range of gradients. During periods of low
sunspot activity (\textit{i.e.}, 2007-2009), neutral lines with and
without filaments  have similar range of gradients (\textit{e.g.}, in
2007, 85\% of neutral lines with and  without filaments were below
about 0.005 G km$^{-1}$ (c.f., Figure 7 left and right panels).

Thus, our data show a clear tendency for NLs with filaments  to have
stronger magnetic gradients during the rising and declining phases and
the maximum of solar cycle, as compared with NLs without filaments.
Although filaments are formed above the NLs with a wide range of
gradients, there appears to be a preferred range of gradients where
filaments are observed more often. This preferred range covers
gradients from about 0.0025 to 0.019 G km$^{-1}$. Less than 20\% of
filaments  were formed at gradients below about 0.001 G km$^{-1}$, and
less than about 5\% of filaments have developed above NLs with 0.0019
G km$^{-1}$ (at solar maximum).  On the other hand, having "preferred"
horizontal gradient does not guarantee the filament formation there;
large fraction of  NLs which exhibit similar gradients remains empty
of filament material.  Furthermore, on average, only a small fraction
of neutral lines is associated with filaments. This fraction varies
from about 2\% at solar minimum to 15\%  at solar maximum (Figure 8).

\begin{figure}
\includegraphics[width=1.0\textwidth,clip=]{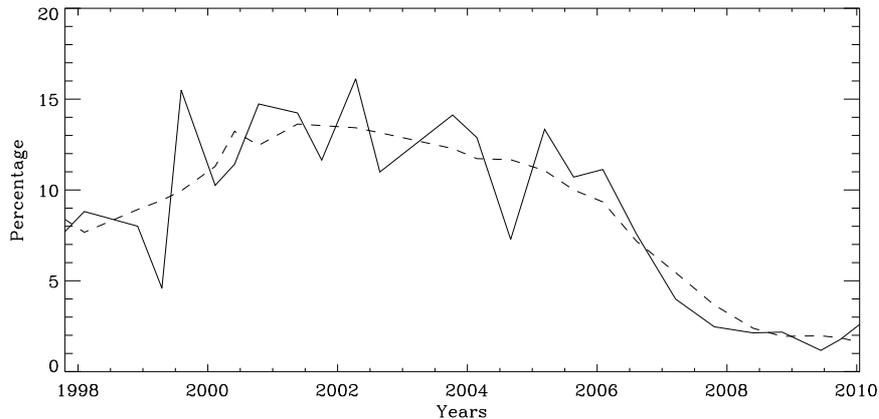}
\caption{Fraction of NL total length occupied by filaments as a function of 
time. Solid line shows measurements from selected Carrington rotations, and 
the dashed line represents 5-point central moving average.
\label{fig8}}
\end{figure}
 
Next, we investigate the properties of magnetic field gradients for
different types of filaments. One of the earliest classification
schemes (\textit{e.g.}, \opencite{Tang1987}) segregates filaments on
two categories based on the nature of the NL above which the filament
lies.  If filament overlays a neutral line separating two polarities
forming an active region, it is classified as a `bipolar region
filament', or type A. When filament is formed above a NL that
separates two different active regions, it is called a `between
bipolar region filament', or type B. To automate the categorization of
NLs, we assigned their type based on magnetic polarity orientation and
the Hale polarity rule (\textit{i.e.}, if the (leading-following)
polarity orientation for a given NL agrees with the Hale-Nicholson
polarity rule, this NL and filament above it were classified as type
A). Otherwise, the NL (and a filament) were classified as type
B. Example of such classification is shown in Figure 2. Green color is
used for type A lines and red color -- for type B. Although this
simplified approach allows an automatic identification of the type of
NLs and filaments, it has a limitation of mis-identifying NLs and
filaments associated with non-Hale polarity regions. Likely, only a
small fraction (about a few percent) of all active regions disobeys
the Hale-Nicholson polarity rule.

Two types of NLs (and filaments) do not appear to show any significant
difference in properties of magnetic gradients. We did find, however,
a slight preference for filaments to form in between active regions
(type B) at a rising phase of solar cycle, while type A filaments
(inside active region) were more abundant at the declining phase of
cycle.

Lastly, we expanded  our investigation to the neutral lines derived
from the chromospheric magnetic fields observed by \textit{Vector
Spectromagnetograph} (VSM) on SOLIS. We applied the same technique as
for the photospheric magnetograms. The results are very similar to
those derived from the photospheric fields with respect to the pattern
of  gradients  described above.
                     
\section{Discussion\label{sec:summary}}
                                   
We studied the properties of horizontal gradients across neutral lines
depending on presence or absence of filaments above them.
We found a preferred range of NL gradients where filaments are
observed relatively more frequently.  Still only about 18\% of neutral
lines within that "preferred range" of gradients have filaments above
them. Thus, a mere fact that a horizontal gradient across magnetic
neutral line falls within a "preferred" range does not guarantee that
a filament will form in that location.  One cannot exclude, however,
that together with other factors these gradient properties could play
a role in creating overall conditions for filament formation. Thus,
for example, several studies suggested  that convergence of magnetic
field may be required for the filament formation (\textit{e.g.},
\opencite{Martin1990}; \opencite{Martin1998}; \opencite{Mackay2008}).
Another factor that could play important role in filament formation is
the rate of canceling magnetic features that were hypothesized to
supply the material to filaments \cite{Martin1998,Litvinenko1999}.
Due to spatial and temporal limitation of the data set used in this
study, we did not investigate the role of these other factors in
filament formation.

The gradients for neutral lines with filaments show similar variations
with the solar cycle as do the neutral lines without filaments
(\textit{i.e.}, an increase in fraction of stronger gradients during
the maximum of  solar activity).  However, this change in properties
of gradients for NLs with filaments is enhanced during the rising and
declining phases of cycle, while the properties of NLs without
filaments are more narrowly defined to a maximum of solar cycle.
During these periods (of rising and declining  phases of solar cycle),
a larger fraction of NLs within preferred range of gradients is
occupied by filaments.

Although we did not find a strong preponderance of filament formation
on the properties of magnetic gradients of neutral lines, our data do
suggest that there is a range of magnetic gradients preferred for
filament formation. A cyclic increase in fraction of neutral lines
(with these preferred gradients) may provide enhancement in favorable
conditions for more filament formation during solar maximum. However,
there is a need for an additional mechanism(s) that takes advantage of
the above favorable conditions to create more filaments. Such
additional condition could be the amount of material in the corona
available for filament formation. One can speculate that during solar
maximum, overall conditions in the photosphere and chromosphere
(\textit{e.g.}, enhanced network fields, larger number of small-scale
reconnection events \textit{etc}.) may result in a significantly larger amount
of material supplied to the corona. In combination with the increased
fraction of NLs with stronger gradients, having more material in the
corona may create more opportunities for condensation of this material
to form the chromospheric filaments.

Finally, the noted displacement of filaments (with respect to the
photospheric neutral lines) and the lack of a strong correlation
between the gradient properties of NLs and filament formation may
suggest that filament formation is governed by properties of coronal
magnetic fields rather than the photospheric or chromospheric magnetic
fields. In itself, this may fit better with the model in which the excess of
material in the corona combined with changes in properties of NLs at
solar cycle maximum is the prime reason for a solar cycle variation
in chromospheric filaments. At the present, no such model exists.

\begin{acks}
The authors acknowledge funding from NASA's NNH09AL04I inter agency
transfer and NSF grant AGS-0837915.  This work utilizes SOLIS data
obtained by  the NSO Integrated Synoptic Program (NISP), managed by
the National  Solar Observatory, which is operated by the Association
of Universities for  Research in Astronomy (AURA), Inc. under a
cooperative agreement with  the National Science Foundation. SOHO is a
project of international cooperation between ESA and NASA.  National
Solar Observatory (NSO) is operated by the Association of Universities
for Research in Astronomy,  AURA Inc under cooperative agreement with
the National Science Foundation (NSF).
\end{acks}

\end{article}
\end{document}